\documentclass{iopart}
\usepackage{graphicx}
\usepackage{color}

\begin{document}
\title{On the Response of Particle Detectors in Vaidya Spacetimes}

\author{A.M.Venditti$^1$, C.C.Dyer$^2$}

\address{Department of Physics, University of Toronto, 50 St. George Street, Toronto, Ontario M5S 3H4, Canada$^1$}
\address{Department of Astronomy and Astrophysics, University of Toronto, 50 St. George Street, Toronto, Ontario M5S 3H4, Canada$^2$}
\address{Department of Physical and Environment Sciences, University of Toronto at Scarborough, 1265 Military Trail, Scarborough, Ontario M1C 1A4, Canada$^2$}
\eads{\mailto{avenditt@physics.utoronto.ca$^1$}, \mailto{dyer@astro.utoronto.ca$^2$}}
\begin{abstract}
Using the formalism of the interaction picture we calculate an expression for the Wightman function for only the spherically symmetric modes of a quantum Klein-Gordon scalar field in a general Vaidya spacetime with ingoing null dust.  It is demonstrated that particle detectors following time-like trajectories that are in the ground state at some time outside of the collapsing shell will respond independently from the configuration of the ingoing null dust if the response is taken at any time outside of the collapsing shell.  For detectors that are taken to be in the ground state at a time interior to the shell it is shown that their response will depend on the configuration of the ingoing null dust.  Relevance to the information loss paradox is discussed.
\end{abstract}
\submitto{\CQG}

\section{Introduction}
In Hawking's original paper on black hole radiation \cite{Hawking_1975}, a general spherically symmetric collapse of matter is considered in which there is no interaction between the quantum field and the classical matter composing the black hole.  The trade-off for considering such a general collapse scenario was that one could not calculate the detailed spectrum of radiation everywhere in the spacetime.\\

Hawking later demonstrated that if the radiation coming from a black hole was exactly thermal then one would lose all information about the matter that formed the black hole \cite{Hawking_1976}.  The lose of information was expressed mathematically, in this case, by the possible non-unitary evolution of quantum states, this is the famous information paradox.\\

There have been several attempts to solve the information lose paradox by demonstrating that the information is not lost at all.  One class of these solutions is to demonstrate that when the back reaction of the emitted radiation is allowed to affect the mass of the black hole then information is not lost at all, for example see \cite{Parikh_Wilczek_2000}, \cite{Visser_et_al_2008}.  There have also been claims that a full theory of quantum gravity is unitary and therefore information preserving \cite{Hawking_2005}.  \\

Other attempts claim simply that if the radiation is calculated in detail for specific spacetimes then one would find the radiation is not thermal and hence information can be carried away in the deviations from thermality \cite{Singh_Vaz_2000}, \cite{Ford_Parker_1978}, \cite{Krauss_2007}.\\

A third class of attempts to resolve the information paradox is to demonstrate that the radiation is exactly thermal and information is lost and to demonstrate that the laws of physics would still be consistent \cite{Unruh_Wald_1995}.\\  

This paper will demonstrate that, in the absence of backreaction in Vaidya spacetimes, a particle detector that follows a time-like trajectory which is coupled to a spherically symmetric, massless Klein-Gordon field will not be able to distinguish between the different null dust configurations that collapse to form the black hole.  Hence, the conclusions of this paper are in line with that of \cite{Unruh_Wald_1995}, \cite{Hawking_1976} and \cite{Hawking_1975} and opposed to the conclusions of \cite{Singh_Vaz_2000}, \cite{Ford_Parker_1978}, \cite{Krauss_2007} with regards to the thermality of the outgoing radiation.\\

In this paper the Vaidya metric will be used as a model for a black hole formed by null dust (see \cite{Poisson_2004}).  
\begin{equation} \label{eq:vaidya}
ds^2 = (1-2m(v)/r)dv^2 - 2dvdr - r^2d\theta^2 - r^2\sin^2(\theta)d\varphi^2
\end{equation}
where $v$ is the ingoing null coordinate that is constant on radially ingoing null trajectories.  We will work with a mass function $m(v)$ that is of the form
\begin{eqnarray} \label{eq:massfunction}
   m(v) = \left\{
     \begin{array}{lr}
       0 &  v < 0\\
       f(v) &  0 \leq v \leq T \\
       m_{0} &  v > T
     \end{array}
   \right.
\end{eqnarray} 
where $f(v)$ is some increasing function that goes from $0$ to $m_{0}$ and $T$ is some timescale.  Physically, the Vaidya class of spacetimes with the above mass function correspond to a spacetime with a spherically symmetric shell with finite thickness of ingoing light collapsing to a point to form a black hole.  Outside of the shell the spacetime is Schwarschild and inside the spacetime is Minkowski.  From now on we will refer to the regions $v < 0$, $0 < v < T$ and $v > T$ as regions $\alpha$, $\beta$ and $\gamma$ respectively.\\

The spherically symmetric modes (s-waves) of a complex, massless scalar field will be quantized in this background.  The Wightman function will be calculated using only the spherically symmetric modes and the response of a particle detector will be evaluated for a constant $r$ observer.  The particle detector response can be thought of as due only to the spherical modes.  It will be shown that the response rate of the particle detector at a point outside of the collapsing shell will not depend on the mass function $m(v)$ if the detector is taken to be in its ground state at a null time outside of the collapsing null dust.  Therefore the information about the matter used to form the black hole cannot be present in correlations in the outgoing radiation.\\

\section{Quantization}
The action for a complex, massless scalar field in the Vaidya background is
\begin{equation}
S = \int{d^4x\sqrt{-g} \partial_{a}\phi\partial_{b}\phi^{*}g^{ab}}
\end{equation}
with $g_{ab}$ taken to be the above metric.  The above action will be only linear in time derivatives of the field.  Therefore, the system is a constrained one and this will have to be taken into account when performing the canonical quantization.  \\

It will be shown that the Hamiltonian for the system can be written in the form
\begin{equation}\label{eq:flatcurvedhamiltonian}
H = H_{0} + H_{1}(v)
\end{equation}
The $H_{0}$ part will be the Hamiltonian for a massless complex scalar field in flat spacetime.  It will be convenient to work in a picture where the states evolve under the $H_{1}(v)$ part of the Hamiltonian while the operators evolve under the $H_{0}$ part.  Thus the scalar field operators will evolve as if they were in flat spacetime and the rest of the time evolution due to the curved geometry will be shunted onto the state.\\

Assuming no angular dependence of the scalar field (s-waves) we can expand the field $\phi$ as follows

\begin{equation}
\phi(v, r) = \sum_{k}a_{k}(v)f_{k}(r)
\end{equation}

Substituting the above expansion in the action and substituting in the Vaidya metric the action reduces to

\begin{equation}
S = \int{dv \sum_{k, l}\left( m(v) a_k^{*}C_{kl}a_l - a_{k}^{*}B_{kl}a_{l} - \rmi a_{k}^{*}A_{kl}\dot{a}_{l}\right) }
\end{equation}
where the $\dot{a}_{k}$ denotes differentiation of $a_{k}$ with respect to the ingoing null time $v$.  The Hermitian matrices $A$, $B$ and $C$ are defined as

\begin{eqnarray} \label{eq:matrices}
\label{eq:amatrix} A_{kl} &=& -4\pi \rmi \int_{0}^{\infty}{dr r^2\left(f_{l}\partial_{r}f^{*}_{k} - \partial_{r}f_{l}f^{*}_{k}\right)} \\
\label{eq:bmatrix} B_{kl} &=& 4\pi\int_{0}^{\infty}{dr r^2 \partial_{r}f_{l}\partial_{r}f^{*}_{k}} \\
\label{eq:cmatrix} C_{kl} &=& 8\pi\int_{0}^{\infty}{dr r \partial_{r}f_{l}\partial_{r}f^{*}_{k}} 
\end{eqnarray}

The action can be re-written in a more convenient form by ignoring a boundary term.  The re-written action is
\begin{eqnarray} \label{eq:lagrangian}
S &=& \int{dv L(v)} \\
\nonumber  &=& \int{dv\left( -\frac{\rmi}{2}a^{\dagger}A\dot{a} + \frac{\rmi}{2}\dot{a}^{\dagger}Aa - a^{\dagger}Ba + m(v)a^{\dagger}Ca\right)}
\end{eqnarray}
where $a$ is a column vector of the modes defined by $a := (a_{1}, a_{2}, ...)^{t}$.  The conjugate momenta is given by the usual formulae
\begin{eqnarray}
p &=& \frac{\delta S}{\delta \dot{a}} = -\frac{\rmi}{2}a^{\dagger}A \\
p^{\dagger} &=& \frac{\delta S}{\delta \dot{a}^{\dagger}} = \frac{\rmi}{2}Aa
\end{eqnarray}
where $p$ is the row vector $p := (p_{1}, p_{2}, ...)$ and $p^{\dagger}$ is the corresponding column vector.  Hence there are two infinite sets of primary constraints between the conjugate momenta ($p_{k}$ and $p^{\dagger}_{k}$) and the $a_{k}$, $a^{\dagger}_{k}$.  In order to quantize this system we follow the Dirac procedure outlined in \cite{Henneaux_Teitelboim_1994}.  We denote the constraints for some conjugate momenta $p_{k}$ from the first set of constraints (\ref{eq:constraint1}) and $p^{\dagger}_{k}$ from the second set of constraints (\ref{eq:constraint2}) respectively by
\begin{eqnarray}
\label{eq:constraint1} G_{1,k} &=& p_{k} + \sum_{l}\frac{\rmi}{2}a^{\dagger}_{l}A_{lk} = 0\\
\label{eq:constraint2} G_{2,k} &=& p^{\dagger}_{k} - \sum_{l}\frac{\rmi}{2}A_{kl}a_{l} = 0
\end{eqnarray}

The canonical Hamiltonian defined by the Legendre transformation is
\begin{equation}
H = p\dot{a} + \dot{a}^{\dagger}p^{\dagger} - L = a^{\dagger}Ba - m(v)a^{\dagger}Ca
\end{equation}
where $L$ is the Lagrangian given by (\ref{eq:lagrangian}).  Using the primary constraints in equation (\ref{eq:constraint1}), (\ref{eq:constraint2}) we can re-write the Hamiltonian as
\begin{eqnarray}\label{eq:constrainthamiltonian}
H &= 2pA^{-1}BA^{-1}p^{\dagger} - 2m(v)pA^{-1}CA^{-1}p^{\dagger}\\
\nonumber  &+ \frac{1}{2}a^{\dagger}Ba - \frac{1}{2}m(v)a^{\dagger}Ca
\end{eqnarray}
We now show that the time derivative of each of the constraints $G_{1,k}$, $G_{2,k}$ is equal to some multiple of themselves.  The time evolution is generated by the Hamiltonian in equation (\ref{eq:constrainthamiltonian}).  Using the normal Poisson bracket relations $\{a_{k},p_{l}\} = \delta_{lk}$ and $\{a^{\dagger}_{k}, p^{\dagger}_{l}\} = \delta_{lk}$ with all other Poisson brackets zero, the following can be shown:
\begin{eqnarray}
\dot{G}_{1} &=& \{G_{1}, H\} = \rmi G_{1}(A^{-1}B - m(v)A^{-1}C) \\
\dot{G}_{2} &=& \{G_{2}, H\} = -\rmi(BA^{-1} - m(v)CA^{-1})G_{2}
\end{eqnarray}
where $G_{1}$ is a row vector and $G_{2}$ is a column vector.  Hence we have a complete and consistent system of constraints and Hamiltonian.\\

In order to quantize this system we observe that
\begin{equation}
\{G_{1}, G_{2}\} = \rmi A \neq 0
\end{equation}
These constraints are called ``second class''.  To quantize the system the Dirac procedure will be used where the commutation relations of the observables are given by
\begin{equation}
[\hat{f},\hat{g}] = \rmi\{f,g\}_{DB}
\end{equation}
where $\hat{f}$ and $\hat{g}$ are the operators corresponding to the classical variables f and g respectively.  $\{f,g\}_{DB}$ is the Dirac bracket given by the following formula 
\begin{equation}
\{f,g\}_{DB} = \{f,g\} - \sum_{a, b}\{f,G_{a}\}F^{ab}\{G_{b}, g\}
\end{equation}
where the $G_{a}$ are all of the second class constraints and $F^{ab}$ is the inverse of the matrix $\{G_{a},G_{b}\}$.\\

Following the same procedure as seen in \cite{Krauss_2007} we use the principal axis transformation (\cite{Goldstein_2001}) to choose the basis $a_{k}$ so that $A$ is the identity matrix and $C$ is some diagonal matrix with real entries denoted by $\lambda_{k}$ for each mode $a_{k}$.  The variables $a_{k}$ are complex so a new basis can be chosen so that the negative eigenvalues of the matrix $A$ become equal to $1$ by re-scaling the modes $a_{k}$ by $\rmi$.  The zero eigenvalues of the matrix $A$ can be ignored as any mode with a corresponding zero eigenvalue for $A$ would have no time derivative term appearing in the action for that mode (\ref{eq:lagrangian}) and would thus be non-dynamical.\\

Now that the $A$ matrix is the identity we can deal with a single set of modes $a_{k}$, $a^{\dagger}_{k}$, $p_{k}$, $p^{\dagger}_{k}$.  The Dirac bracket is readily calculated between each pair of these variables and the commutation relations are given by the rule in (19).  The commutation relations are 
\begin{eqnarray}
\left[ \hat{a}_{k}, \hat{p}_{l} \right] &=& \frac{\rmi}{2}\delta_{kl} \\
\left[ \hat{a}_{k}, \hat{p}^{\dagger}_{l} \right] &=& 0 \\
\left[ \hat{a}_{k}, \hat{a}^{\dagger}_{l} \right] &=& -\delta_{kl} \\
\left[ \hat{a}^{\dagger}_{k}, \hat{p}_{l} \right] &=& 0 \\
\left[ \hat{a}^{\dagger}_{k}, \hat{p}^{\dagger}_{l} \right] &=& \frac{\rmi}{2}\delta_{kl} \\
\left[ \hat{p}_{k}, \hat{p}^{\dagger}_{l} \right] &=& \frac{1}{4}\delta_{kl}
\end{eqnarray}

The Hamiltonian in terms of the new basis which diagonalizes $A$ and $C$ is given as
\begin{eqnarray}
\hat{H} &= \sum_{k, l}2\hat{p}_{l}B_{lk}\hat{p}^{\dagger}_{k} + \frac{1}{2}\hat{a}^{\dagger}_{k}B_{kl}\hat{a}_{l} \\
\nonumber  &- 2m(v)\lambda_{k}\hat{p}_{k}\hat{p}^{\dagger}_{k} - \frac{1}{2}m(v)\hat{a}^{\dagger}_{k}\hat{a}_{k}\lambda_{k}
\end{eqnarray}
Now we divide this Hamiltonian up into the flat spacetime part and the curved spacetime part as in equation (\ref{eq:flatcurvedhamiltonian}) with 
\begin{eqnarray}
\hat{H}_{0} &=& \sum_{k,l} 2\hat{p}_{l}B_{lk}\hat{p}^{\dagger}_{k} + \frac{1}{2}\hat{a}^{\dagger}_{k}B_{kl}\hat{a}_{l} \\
\hat{H}_{1} &=& \sum_{k,l}- 2m(v)\lambda_{k}\hat{p}_{k}\hat{p}^{\dagger}_{k} - \frac{1}{2}m(v)\hat{a}^{\dagger}_{k}\hat{a}_{k}\lambda_{k}
\end{eqnarray}
To work in the interaction picture we demand that the operators become time dependent and evolve under $\hat{H}_{0}$.  Using the commutation relations obtained above we solve the Heisenberg equations for the fundamental operators.
\begin{eqnarray}
\rmi\dot{\hat{p}}_{k} &=& [\hat{p}_{k}, \hat{H}_{0}] = \sum_{l}\left(\frac{1}{2}\hat{p}_{l} - \frac{\rmi}{4}\hat{a}^{\dagger}_{l}\right)B_{lk} \\
\rmi\dot{\hat{a}}_{k} &=& [\hat{a}_{k}, \hat{H}_{0}] = \sum_{l}B_{kl}\left(\rmi \hat{p}^{\dagger}_{l} - \frac{1}{2}\hat{a}_{l}\right) \\
\rmi\dot{\hat{p}}^{\dagger}_{k} &=& [\hat{p}^{\dagger}_{k}, \hat{H}_{0}] = \sum_{l}B_{kl}\left(-\frac{1}{2}\hat{p}^{\dagger}_{l} -\frac{\rmi}{4}\hat{a}_{l}\right) \\
\rmi\dot{\hat{a}}^{\dagger}_{k} &=& [\hat{a}^{\dagger}_{k}, \hat{H}_{0}] = \sum_{l}\left(\rmi \hat{p}_{l}+\frac{1}{2}\hat{a}^{\dagger}_{l}\right)B_{lk}
\end{eqnarray}
Taking linear combinations of the above equations they can be solved relatively easily.  The general solution is
\begin{eqnarray} 
\label{eq:operatorsolutions} \hat{a} &=& -2\rmi e^{\rmi Bv}\hat{J}^{\dagger} + 2i\hat{D}^{\dagger} \\
\hat{p} &=& \hat{J}e^{-\rmi Bv} + \hat{D} 
\end{eqnarray}
where $\hat{J}^{\dagger}$ and $\hat{D}^{\dagger}$ are column vectors of operator integration constants.  To be clear $\hat{J}^{\dagger} = (\hat{j}^{\dagger}_{1}, \hat{j}^{\dagger}_{2}, ...)^t$ so that the $\dagger$ acting on the vector of operators also acts on each individual operator.\\

Imposing the commutation relations between the $\hat{a}$, $\hat{a}^{\dagger}$, $\hat{p}$ and $\hat{p}^{\dagger}$'s we find that the only noncommuting pair of variables from the set $(\hat{J}_{a}, \hat{J}^{\dagger}_{a}, \hat{D}_{a}, \hat{D}^{\dagger}_{a})$ are $\hat{J}_{a}$ and $\hat{J}^{\dagger}_{a}$. The following commutation relations are found
\begin{equation}
\left[ \hat{J}_{a}, \hat{J}^{\dagger}_{b} \right] = \frac{1}{4}\delta_{ab}
\end{equation}
In particular the operators $\hat{D}_{a}$ and $\hat{D}^{\dagger}_{a}$ commute with everything.  From now on we choose $\hat{D} = 0$.  With this choice the flat spacetime Hamiltonian $\hat{H}_{0}$ will not depend on time ($v$), as it would if $\hat{D} \neq 0$, which is a physically reasonable requirement.  Further, the choice $\hat{D} = 0$ is necessary to have the same relations between the operators $\hat{p}$ and $\hat{a}$ as in the classical relations (\ref{eq:constraint1}) and (\ref{eq:constraint2}) in a basis where the matrix $A$ is the identity.\\

In terms of the above solutions to the Heisenberg equations we can re-write $\hat{H}_{0}$ as
\begin{equation}
\hat{H}_{0} = \sum_{k,l}4\hat{J}_{l}B_{lk}\hat{J}^{\dagger}_{k}
\end{equation}
Since $B_{lk}$ is hermitian it can be diagonalized by performing a unitary transformation on the operators $\hat{J}_{k}$.  To this end we re-define the operators as $\hat{J}^{\dagger} = U^{\dagger}\hat{\chi}^{\dagger}/2$, where $U^{\dagger}$ is unitary.  The flat spacetime Hamiltonian becomes
\begin{equation}
\hat{H}_{0} = \sum_{k}\omega_{k}\left(\hat{\chi}^{\dagger}_{k}\hat{\chi}_{k} + 1\right)
\end{equation}
where the $\omega_{k}$ are the real eigenvalues of the matrix $B$ and $[\hat{\chi}_{l},\hat{\chi}^{\dagger}_{k}] = \delta_{lk}$ holding.  It will be shown in section (\ref{sec:response}) that $\omega_{k} > 0$.  Therefore, the $\hat{\chi}_{k}$ operators are the annihilation operators of positive energy particles with their adjoint being the creation operators.  The ``curved" or time-dependent part of the Hamiltonian becomes
\begin{equation} \label{eq:interactinghamiltonian}
\hat{H}_{1} = -m(v)\hat{\chi}^{\dagger}\exp(\rmi\Omega v) \tilde{\Lambda} \exp(-\rmi\Omega v)\hat{\chi} - m(v)Tr(\tilde{\Lambda})
\end{equation}
where here $\tilde{\Lambda}$ is the C matrix in equation (\ref{eq:cmatrix}) in the new basis which diagonalizes the matrix B.  $\Omega$ is a diagonal matrix with the real diagonal entries being the $\omega_{k}$ that are in the $\hat{H}_{0}$ part of the Hamiltonian.  Specifically we have
\begin{eqnarray}
\Omega &:=& UBU^{\dagger} = diag[\omega1, \omega2, ...] \\
\tilde{\Lambda} &:=& (UCU^{\dagger})^{t}
\end{eqnarray}

The state of the Klein-Gordon field will be taken to be given by
\begin{equation} \label{eq:vacuum}
\hat{\chi}_{k}|0\!\!> = 0
\end{equation}
Hence $|0\!\!>$ is the vacuum for the operators $\chi_{k}$ and $\chi_{k}^{\dagger}$.  It will be shown below, when we calculate the $\hat{\phi}$ field explicitly, that this state is the unique state such that the particle detectors of constant $r$ observers will have a null response when $m(v) = 0$.  For the mass function (\ref{eq:massfunction}) the state has this interpretation in region $\alpha$.  It is the most natural choice of vacuum state for spacetimes of the form (\ref{eq:vaidya}) with (\ref{eq:massfunction}) holding because in region $\alpha$ the observers are in a Minkowski spacetime and completely causally disconnected from the non vacuum region of spacetime (see figure 1).\\

In the picture we are working in the state of the Klein-Gordon field evolves as
\begin{equation} \label{eq:timeevolution}
|\Psi(v)> = \mathcal{T}\exp(-\rmi\int^{v}_{v_{0}}{\hat{H}_{1}(v')dv'})|\Psi(v_{0})>
\end{equation}
where $\mathcal{T}$ is the time ordering operator.  If we pick the initial state to be the vacuum with respect to the above creation and annihilation operators (i.e. $\hat{\chi}_{k}|\Psi(v_{0})> = 0$) then we can see from the form of $\hat{H}_{1}$ that the state evolves in time only by the phase factor $\exp(\rmi\int^{v}_{v_{0}}{m(v')Tr(\tilde{\Lambda})})$ which means that the vacuum state vector is mathematically the same at all times.  Notice however that a general state from the Fock space formed by the creation operators will not evolve only by a phase factor and hence there will be a change in the state due to the Vaidya background.  This can be seen by expanding equation (\ref{eq:timeevolution}) to first order to obtain
\begin{equation}
|\Psi(v)> \approx |\Psi(v_{0})> - \rmi\int^{v}_{v_{0}}{\hat{H}_{1}(v')dv'}|\Psi(v_{0})>
\end{equation}
So now if we have a one particle state $|\Psi(v_{0})> = \hat{\chi}^{\dagger}_{k}|0>$ at time $v=v_{0}$ then to first order we will have a linear combination of one particle states given by
\begin{eqnarray}
|\Psi(v)> \approx &\left(1+\rmi\int_{v_{0}}^{v}{dv' m(v') Tr(\tilde{\Lambda})} \right)|\Psi(v_{0})> + \\
\nonumber &\sum_{p,j,l}\rmi\int^{v}_{v_{0}}{dv'm(v')\exp(\rmi\Omega v')_{lp}\tilde{\Lambda}_{pj}\exp(-\rmi\Omega v')_{jk}}\hat{\chi}^{\dagger}_{l}|0>
\end{eqnarray}

\section{Response of particle detectors}\label{sec:response}

The response rate of a particle detector at a given frequency $\nu$ as it follows some path through spacetime is given by
\begin{equation} \label{eq:responserate}
\dot{F}_{\tau}(\nu) = 2\int_{0}^{\tau-\tau_{0}}{ds \,\,\, \Re(\,\, \exp(-\rmi\nu s)G^{+}(x(\tau), x(\tau - s)))}
\end{equation}
where $x(\tau)$ is the path through spacetime (see \cite{Schlicht_2003}, \cite{Birrell_Davies_1982}) and $\Re$ denotes the real part.  $\tau$ is a proper time of the detector that parameterizes the path $x(\tau)$.  Also, the particle detector is in the ground state at $\tau_{0}$.  First, we must evaluate the Wightman function $G^{+}(x(\tau),x(\tau-s)) = <0|\hat{\phi}(x(\tau))\hat{\phi}(x(\tau-s))|0>$. \\

The scalar field operators are given by
\begin{equation}
\hat{\phi}(v,r) = \sum_{k}\hat{a}_{k}(v)f_{k}(r)
\end{equation}
We have solved for the operator coefficients $\hat{a}_{k}$ in equation (\ref{eq:operatorsolutions}).  The $\hat{\phi}(v,r)$ field operators are then given by
\begin{equation}
\hat{\phi}(v,r) = \sum_{l,a}-2\rmi f_{l}(r)\exp(\rmi Bv)_{la}\hat{J}^{\dagger}_{a}
\end{equation}
where $\exp(iBv)$ is a matrix.  If we now make the substitution $\hat{J}^{\dagger}_{a} = 1/2U^{*}_{ba}\hat{\chi}^{\dagger}_{b}$, which diagonalizes the Hamiltonian $\hat{H}_{0}$, we end up with the field operator as
\begin{equation}
\hat{\phi}(v,r) = \sum_{b, c, l}-\rmi f_{l}(r)U^{*}_{cl}\exp(\rmi\Omega v)_{cb}\hat{\chi}^{\dagger}_{b}
\end{equation}
 Defining $g_{c}(r) = f_{l}U^{*}_{cl}$ and substituting this into the above expression, we finally obtain.
\begin{equation} \label{eq:phiexpansion}
\hat{\phi}(v,r) = \sum_{k}-\rmi g_{k}(r)\exp(\rmi\omega_k v)\hat{\chi}^{\dagger}_{k}
\end{equation}
In order to solve explicitly for the $g_{k}(r)$ modes we use the fact that the complex field $\phi(v,r)$ is a solution of the wave equation determined by the $H_{0}$ operator.  It is not difficult to see that this will be the wave equation in the Vaidya background with $m(v)=0$, i.e. the wave equation in Minkowski spacetime.  By substituting the expansion (\ref{eq:phiexpansion}) into the wave equation for the Vaidya background with $m(v)=0$ we obtain that the equation 
\begin{equation}
r^2\partial_{r}^{2}g_{k}(r) + 2\rmi\omega_k r^2\partial_{r} g_{k}(r) + 2r\partial_{r}g_{k}(r) + 2\rmi\omega_k rg_{k}(r) = 0
\end{equation}
must hold for each mode $g_{k}(r)$.  The general solution to this equation is
\begin{equation} \label{eq:gmodes}
g_{k}(r) = \frac{C(k)}{r} + D(k)\frac{\exp(-2\rmi\omega_k r)}{r}
\end{equation}
where $C(k)$ and $D(k)$ are arbitrary complex constants.  We wish the modes we expand in to be physical and hence regular at $r=0$ in Minkowski spacetime.  The only way to achieve this is to choose $C(k) = -D(k)$ so that the divergence there cancels.\\

The expression for the matrix $B$ is given by (\ref{eq:bmatrix}) in terms of the original $f_{k}(r)$ modes.  The diagonalized matrix $\Omega$ is given in terms of $B$ as
\begin{equation}
\Omega_{lk} = \sum_{a, b}U_{la}B_{ab}U^{*}_{kb} 
\end{equation}
Using the definition of the matrix $B$ and the definition of the modes $g_{k}(r)$ we obtain that 
\begin{equation} \label{eq:omegamatrix}
\Omega_{lk} = 4\pi\int_{0}^{\infty}{dr r^2 \partial_{r} g_{l}^{*}(r) \partial_{r} g_{k}(r)}
\end{equation}
As a check, the modes (\ref{eq:gmodes}) with $C(k) = -D(k)$, when substituted into the right hand side of (\ref{eq:omegamatrix}) should be diagonal in $l$ and $k$.  Substitution of the modes
\begin{equation}\label{eq:gwavemodes}
g_{k}(r) = D(k)(\frac{-1}{r} + \frac{\exp(-2\rmi\omega_k r)}{r})
\end{equation}
into this integral can be shown to reduce to the single term
\begin{eqnarray} \label{eq:checkintegral}
\Omega_{lk} = 16\pi D^{*}(l)D(k)\int_{0}^{\infty}{dr \; \omega_l\omega_k \exp(2i(\omega_l-\omega_k)r) }
\end{eqnarray}
where $\omega_l$ and $\omega_k$ are real numbers.  To show this matrix is diagonal we will put our system in a box of finite size so that we can index our momentum modes with integers.  The radius of our system will be $R$ and this will be the upper bound of the integral in (\ref{eq:checkintegral}).  The discrete limit is obtained by making the replacements
\begin{equation} \label{eq:contrelations}
\omega_l,\omega_k  \;\; \rightarrow \;\; \frac{\pi l}{R}, \frac{\pi k}{R}
\end{equation}
where on the right hand side $l$ and $k$ are integers.  The integral (\ref{eq:checkintegral}) becomes
\begin{equation}
16\pi D^{*}(l)D(k)\int_{0}^{R}{dr \; \frac{\pi^2}{R^2} lk \left[ \cos(\frac{2\pi}{R}(l-k)r) + \rmi\sin(\frac{2\pi}{R}(l-k)r) \right] }
\end{equation}
This is easily shown to be equal to
\begin{equation} \label{eq:omegaexpression}
\frac{16\pi^3}{R}l^2\delta_{lk}|D(l)|^2
\end{equation}
Recall that the elements of $\Omega_{lk}$ are $\omega_k$ on the diagonal.  Using this in the discrete limit we have that the above expression must be equal to $\pi l/R$, in the discrete limit.  This fixes the the magnitude of the complex constant to be 
\begin{equation} \label{eq:gconstant}
|D(l)| = \frac{1}{4\pi\sqrt{l}}
\end{equation}
The expression (\ref{eq:omegaexpression}) is manifestly positive.  Therefore, only the modes $\omega_k > 0$ need to be considered when we calculate the Wightman function.\\

The quantized complex Klein-Gordon field in (\ref{eq:phiexpansion}) is not Hermitian and therefore not a valid observable.  We will use the Hermitian observable that is formed from equation (\ref{eq:phiexpansion}) which is given by the following expansion
\begin{equation} \label{eq:realscalarfield}
\Re\left( \hat{\phi}(v,r) \right) = \sum_{k}\frac{-\rmi}{2}(g_{k}(r)\exp(\rmi\omega_k v)\hat{\chi}^{\dagger}_{k} - g^{*}_{k}(r)\exp(-\rmi\omega_k v)\hat{\chi}_{k})
\end{equation} 
with the $g_{k}$ given by equation (\ref{eq:gwavemodes}) and (\ref{eq:gconstant}).  The above Hermitian field obeys the correct equations of motion and the correct commutation relations and is therefore the real, quantum, spherically-symmetric Klein-Gordon field operator.\\

In order to calculate the Wightman function using the interaction picture we will first calculate the Feynman correlation function using the formula given in \cite{Srednicki_2007}.
\begin{eqnarray}\label{eq:interactioncor}
&<\oslash|\mathcal{T}\hat{\phi}_{F}(v_{n},r_{n})\dots\hat{\phi}_{F}(v_{1},r_{1})|\oslash> \\
\nonumber &= \frac{<0|\mathcal{T}\hat{\phi}(v_{n},r_{n})\dots \hat{\phi}(v_{1},r_{1})\exp(-\rmi\int_{-\infty}^{\infty}{dv \hat{H}_{1}(v)})|0>}{<0|\mathcal{T}\exp(-\rmi\int_{-\infty}^{\infty}{dv \hat{H}_{1}(v)})|0>}
\end{eqnarray}
The state $|\oslash>$ signifies the vacuum state of the full Hamiltonian $\hat{H} = \hat{H}_{0}+\hat{H}_{1}(v)$ and $|0>$ is the vacuum state of the ``free'' Hamiltonian $\hat{H}_{0}$ and $\hat{\phi}_{F}(v,r)$ denote the scalar field with the full Hamiltonian.  Note that the contribution from the integrals at $\infty$ cancel since $\hat{H}_{1}(v)=0$ for $v < 0$ (in region $\alpha$).  Also, the $Tr(\tilde{\Lambda})$ term in $\hat{H}_{1}(v)$ cancels.  For the case of two $\hat{\phi}$ factors the expression becomes for $v_{2} > v_{1}$
\begin{eqnarray}\label{eq:twopointcor}
&<\oslash|\hat{\phi}_{F}(v_{2},r_{2})\hat{\phi}_{F}(v_{1},r_{1})|\oslash> \\
\nonumber &= <0|\hat{\phi}(v_{2},r_{2})\mathcal{T}\exp(-\rmi\int_{v_{1}}^{v_{2}}{dv \eta(v)}) \hat{\phi}(v_{1},r_{1})|0>\\
\nonumber &\eta(v) \equiv -m(v)\hat{\chi}^{\dagger}\exp(\rmi\Omega v) \tilde{\Lambda} \exp(-\rmi\Omega v)\hat{\chi}
\end{eqnarray} 
where we have used the definition of the vacuum state (\ref{eq:vacuum}) to evaluate the denominator in (\ref{eq:interactioncor}).\\

At this point we can specify the meaning of the vacuum state given in (\ref{eq:vacuum}).  Using the general formula for the Wightman function (\ref{eq:interactioncor}) we can obtain the Wightman function in Minkowski spacetime for the spherically symmetric modes of the real scalar field operator given in (\ref{eq:realscalarfield}).  In the continuum limit it is expressed by the integral
\begin{equation}\label{eq:MinkowskiWightman}
G^{+}(t, r; t', r') = \frac{1}{16\pi r r'}\int_{0}^{\infty}{\frac{d\omega}{\omega} \exp\left(-\rmi\omega(t - t')\right)\sin(\omega r)\sin(\omega r')}
\end{equation}
where we have used that $v = t+r$ in Minkowski spacetime and we have used the relations (\ref{eq:contrelations}) and the relation
\begin{equation}
\frac{1}{R}\sum_{k} \rightarrow \int{d\omega}
\end{equation}
to go to the continuum limit.  By plugging (\ref{eq:MinkowskiWightman}) into equation (\ref{eq:responserate}) it can be verified that constant $r$ observers in Minkowski spacetime in the state (\ref{eq:vacuum}) will have a null response for their particle detectors (see appendix for details).\\

We note that the $v=const$ surfaces are not Cauchy and thus there is a concern that the field evolution will not be unique if we specify the evolution in terms of these surfaces.  For Vaidya metrics of the form (\ref{eq:vaidya}), (\ref{eq:massfunction}) we have that for constant $v$ as $r \rightarrow \infty$ the metric becomes Minkowski spacetime.  The limit $r \rightarrow \infty$ can effectively be reproduced by considering $m(v) = 0$.  Therefore an alternate characterization of the vacuum (\ref{eq:vacuum}) is that it is the vacuum state such that constant $r$ particle detectors do not click on past null infinity (see figure \ref{fig:figure2}).  Past null infinity is a Cauchy surface and thus specifying the state of the field there is sufficient to determine the evolution of the field uniquely throughout the spacetime.\\

Another way to see that the evolution of the Klein-Gordon field is unique is to note that imposing the condition that $\hat{\phi}$ be finite at $r=0$ as well as picking the modes to be positive frequency with respect to the inertial time $t$ in Minkowski spacetime fixes all arbitrary constants present in the mode solutions to the Klein-Gordon equation.\\

The proper time of the constant $r$ observers will be used to parametrize the paths in formula (\ref{eq:responserate}).  In order to evaluate the response rate of a particle detector for a constant $r$ observer it is convenient to re-write the equation (\ref{eq:responserate}) in terms of an integral over $v$ instead of the proper time of the time-like particle detector.  To this end, make the substitution $p = \tau - s$ so that we now integrate over the variable $p$ instead of $s$.  $\tau$ is a constant representing the proper time at which the response of the detector is evaluated at.
\begin{equation}
\dot{F}_{\tau}(\nu) = 2\int_{\tau_{0}}^{\tau}{dp\Re\left(\exp(-\rmi \nu (\tau - p))\right) G^{+}(x(\tau), x(p))}
\end{equation}
Since $\tau$ and $p$ are both proper times we can use the line element (\ref{eq:vaidya}) to solve for them in terms of the null time $v$.  
\begin{equation} \label{eq:propertime}
d\tau = \sqrt{1-\frac{2m(v)}{r}}dv
\end{equation}
The proper times as functions of the null time are denoted as $\tau = \tau(v)$ and $p = \tau(v')$.  In the integral over the proper time it is $p$ that is integrated over, so in terms of null time we will be integrating over $v'$.  The final formula is then given by
\begin{eqnarray} \label{eq:responserate2}
&\dot{F}_{v}(\nu) = 2\int_{v_{0}}^{v}{dv' \,\,\, \Re(\xi(v',r)G^{+}(v,r; v', r))} \\
\nonumber &\xi(v',r) \equiv \sqrt{1-\frac{2m(v')}{r}} \exp(-\rmi\nu (\tau(v)-\tau(v')))
\end{eqnarray}
where $G^{+}(v,r;v',r)$ is given by equation (\ref{eq:twopointcor}) and the detector is in the ground state at $v_{0}$.  The above formula represents the transition rate for an observer at radius $r$ at time $v$.\\

For mass functions of the form (\ref{eq:massfunction}) it is clear from the form of the Wightman function given in (\ref{eq:twopointcor}) that for $v_{1}$ and $v_{2}$ in region $\gamma$ it will be independent of the mass function in region $\beta$.  Therefore, the Wightman function is completely independent of the details of the collapse to form a black hole when evaluated in this region.  Hence, from equation (\ref{eq:responserate2}) we have that the transition rate of a detector that is in its ground state at some time in region $\gamma$ is completely independent of the details of the collapse.\\

It is clear from equation (\ref{eq:twopointcor}) that the Wightman function will depend on the mass function if evaluated at null times such that $v_{1}$ is in region $\alpha$ and $v_{2}$ is in region $\gamma$ or vice versa.  From equation (\ref{eq:responserate}) it is seen that this indicates that the transition rate of a detector at a time $v$ in region $\gamma$, that is in its ground state at some time $v_{0}$ in region $\alpha$, will depend on the details of the mass function.  This is expected since the detector is actually in the non vacuum region of the spacetime in this scenario (see figure 1).\\

\begin{figure}[htbp]
\begin{center}
 
\input{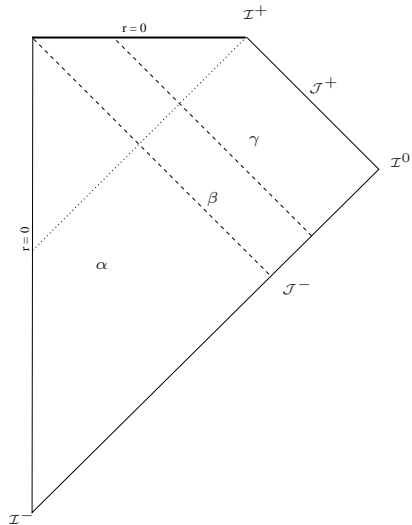}
 
\caption{\small Conformal diagram for a Vaidya spacetime.  The region in between $v = 0$ and $v = T$ is the non vacuum region.  Any massive observer with finite energy (that cannot approach the speed of light) starts on $\mathcal{I}^{-}$ and can end either in the singularity or at $\mathcal{I}^{+}$; hence the observer must travel through the non vacuum region.}
\label{fig:figure1}
\end{center}
\end{figure}


\section{Discussion}
A particle detector that responds only to the spherically symmetric modes of the quantum Klein-Gordon field has been considered in the Vaidya class of spacetimes with a mass function given by equation (\ref{eq:massfunction}).  There are two important cases to be considered for a particle detector that follows a time-like path through spacetime.\\

The first is when the particle detector is taken to be in its ground state at a time $v$ in region $\alpha$.  The response rate of this detector at a time outside of the collapsing null dust (region $\gamma$) is seen to depend on the details of mass function and hence the details of the collapse.  This is expected since a time-like particle detector must travel through the non vacuum region of the spacetime and hence ``feel'' the gravitational field caused by the collapsing matter (see figure \ref{fig:figure1}).\\

The second case is when the particle detector is taken to be in its ground state at some time after the collapse (region $\gamma$).  The response rate of the detector at any time in region $\gamma$ is seen to only depend on the value of the mass function in region $\gamma$.  The response rate is independent of the form of the mass function in region $\beta$ and hence is independent of the details of the collapse.  The implication is that any radiation emitted from the black hole is totally independent of the configuration of the null dust that forms the black hole.  Therefore any deviations from a thermal spectrum in the outgoing radiation could not carry information about the matter that collapsed to form the black hole.  This is an unexpected result and is in contradiction to more recent claims that radiation being emitted from a black hole would deviate from thermality and hence might contain details of the configuration of stress-energy that formed the black hole \cite{Krauss_2007}.  Of course the investigation in \cite{Krauss_2007} has to do with a black hole formed from massive matter and not null dust.  However, on the point that general black holes emit non-thermal radiation that transmits information about the collapsing matter, this paper has conclusions opposite to \cite{Krauss_2007}.  The arguments here are in line with claims made in \cite{Hawking_1975} and \cite{Hawking_1976}.\\

We emphasize that we have not calculated the explicit, numerical response of the particle detector.  We have demonstrated that whatever the response is, it must only depend on the final value of the mass function (\ref{eq:massfunction}) in the Vaidya spacetime and be independent of the form of the function (\ref{eq:massfunction}) which physically means that the response is independent of the null dust configuration.  Therefore we cannot verify the response to be thermal as in \cite{Takagi_1986}, \cite{Christensen_Fulling_1977}.\\

\begin{figure}[htbp]
\begin{center}
 
\input{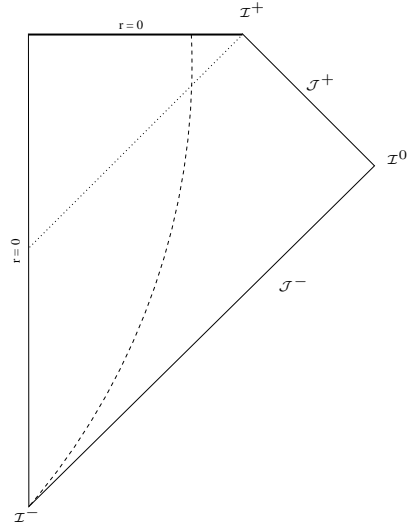}
 
\caption{\small Conformal diagram of massive dust collapsing to form a black hole.  The region behind the dashed curve is the non-vacuum region.  It is easily seen that a massive observer can start at $\mathcal{I}^{-}$ and end at $\mathcal{I}^{+}$ while remaining outside the non vacuum region.}
\label{fig:figure2}
\end{center}
\end{figure}

It is interesting to consider if the results obtained here carry over to the case of a black hole formed by time-like (rather than null) dust (see figure \ref{fig:figure2}).  Specifically the second scenario described above where we consider a detector that is in its ground state when it is outside of the collapsing matter.  In the case of a massive dust collapse, a detector following a time-like path could remain outside of the collapsing matter for all time.  If the response of these detector were in line with the Vaidya case, i.e. their responses do not depend on the configuration of the collapsing matter, then this would again have the implication that the outgoing radiation is totally independent of the collapsing matter.

\ack
We wish to acknowledge the support of an NSERC discovery grant to C. C. Dyer.\\

\section*{Appendix}
We demonstrate that the response of a constant $r$ particle detector in Minkowski space is null in the state with Wightman function given by (\ref{eq:MinkowskiWightman}).  We perform the integral in (\ref{eq:MinkowskiWightman}) by inserting the usual exponential cut-off ($\exp(-\epsilon\omega)$), the result is
\begin{eqnarray}\label{eq:integral}
\int_{0}^{\infty}\frac{d\omega}{\omega}&& \exp\left(-\rmi\omega(t - t')-\epsilon\omega\right)\sin(\omega r)\sin(\omega r') \\
\nonumber &&= \ln\left(\frac{-4r^2+(v-v'+\rmi \epsilon)^2}{(v-v'-\rmi \epsilon)^2}\right)
\end{eqnarray}
To evaluate the response of the detector we plug (\ref{eq:integral}) into (\ref{eq:MinkowskiWightman}) and plug the resulting equation into (\ref{eq:responserate2}).  The resulting expression for the response rate of constant $r$ observers is
\begin{eqnarray}
\frac{1}{16\pi r^2}&&\int^{v}_{-\infty}{dv' \Re \bigg( \exp\left(-\rmi \omega (v - v')\right) \ln\left(-2r + v - v' - \rmi \epsilon \right)} \\
\nonumber &+& \exp\left(-\rmi \omega (v -v')\right) \ln\left(-2r - (v - v') - \rmi \epsilon\right) \\
\nonumber &-& 2 \exp\left(-\rmi \omega (v-v')\right) \ln\left(v - v' - \rmi \epsilon\right)\bigg)
\end{eqnarray}
The above integral can be evaluated numerically to be zero for various values of $\omega$ and $r$ and for small values of $\epsilon$.  It has thus been demonstrated that the vacuum state used in (\ref{eq:vacuum}) is one in which there is a null response for constant $r$ detectors.\\

\bibliographystyle{unsrt}
\bibliography{myrefs}

\begin{thebibliography}{10}

\bibitem{Hawking_1975}
S.~W. Hawking.
\newblock Particle creation by black holes.
\newblock {\em Communications in Mathematical Physics}, 43:199--220, 1975.

\bibitem{Hawking_1976}
S.~W. Hawking.
\newblock Breakdown of predictability in gravitational collapse.
\newblock {\em Physical Review D}, 14(10):2460--2473, 1976.

\bibitem{Parikh_Wilczek_2000}
Maulik~K. Parikh and Frank Wilczek.
\newblock Hawking radiation as tunneling.
\newblock {\em Physical Review Letters}, 85(24):5042--5045, 2000.

\bibitem{Visser_et_al_2008}
Sebastiano~Sonego Carlos~Barcel\'{o}, Stefano~Liberati and Matt Visser.
\newblock Fate of gravitational collapse in semiclassical gravity.
\newblock {\em Physical Review D}, 77, 2008.

\bibitem{Hawking_2005}
S.~W. Hawking.
\newblock Information loss in black holes.
\newblock {\em Physical Review D}, 72, 2005.

\bibitem{Singh_Vaz_2000}
T.~P. Singh and Cenalo Vaz.
\newblock Radiation flux and spectrum in the vaidya collapse model.
\newblock {\em Physics Letters B}, 481:74--78, 2000.

\bibitem{Ford_Parker_1978}
L.~H. Ford and Leonard Parker.
\newblock Creation of particles by singularities in asymptotically flat
  spacetimes.
\newblock {\em Physical Review D}, 17(6):1485--1496, 1978.

\bibitem{Krauss_2007}
Dejan~Stojkovic Tanmay~Vachaspati and Lawrence~M. Krauss.
\newblock Observation of incipient black holes and the information loss
  problem.
\newblock {\em Physical Review D}, 76(2), 2007.

\bibitem{Unruh_Wald_1995}
William~G. Unruh and Robert~M. Wald.
\newblock Evoluation laws taking pure states to mixed states in quantum field
  theory.
\newblock {\em Physical Review D}, 52(4), 1995.

\bibitem{Poisson_2004}
E.~Poisson.
\newblock {\em A Relativist's Toolkit}.
\newblock Cambridge University Press, 2004.

\bibitem{Henneaux_Teitelboim_1994}
Marc Henneaux and Claudio Teitelboim.
\newblock {\em Quantization of Gauge Systems}.
\newblock Princeton University Press, 1994.

\bibitem{Goldstein_2001}
Charles P.~Poole John L.~Safko, Herbert~Goldstein.
\newblock {\em Classical Mechanics}.
\newblock Addison Wesley, 2001.

\bibitem{Schlicht_2003}
Sebastian Schlicht.
\newblock Considerations on the unruh effect: Causality and regularization.
\newblock {\em Classical and Quantum Gravity}, 21:4647--4660, 2004.

\bibitem{Birrell_Davies_1982}
N.D. Birrell and P.C.W. Davies.
\newblock {\em Quantum fields in curved space}.
\newblock Cambridge University Press, 1982.

\bibitem{Srednicki_2007}
M.~Srednicki.
\newblock {\em Quantum Field Theory}.
\newblock Cambridge University Press, 2007.

\bibitem{Takagi_1986}
Shin Takagi.
\newblock Vacuum noise and stress induced by uniform acceleration —
  hawking-unruh effect in rindler manifold of arbitrary dimension —.
\newblock {\em Prog. Theor. Phys. Supplement}, pages 1--142, 1986.

\bibitem{Christensen_Fulling_1977}
S.~M. Christensen and S.~A. Fulling.
\newblock Trace anomalies and the hawking effect.
\newblock {\em Physical Review D}, 15(8), 1977.

\end{thebibliography}

\end{document}